\documentclass[amsmath, amsfonts, twocolumn, prl]{revtex4}
\usepackage{graphicx}
\usepackage{epsfig}
\usepackage{bbm}
\usepackage{bm}
\usepackage{dcolumn}
\usepackage{amsmath}
\usepackage{amssymb}
\usepackage{color}

\def\Xint#1{\mathchoice
   {\XXint\displaystyle\textstyle{#1}}%
   {\XXint\textstyle\scriptstyle{#1}}%
   {\XXint\scriptstyle\scriptscriptstyle{#1}}%
   {\XXint\scriptscriptstyle\scriptscriptstyle{#1}}%
   \!\int}
\def\XXint#1#2#3{{\setbox0=\hbox{$#1{#2#3}{\int}$}
     \vcenter{\hbox{$#2#3$}}\kern-.5\wd0}}

\def\dashint{\Xint-}

\newcommand{\beg}{\begin{equation}}
\newcommand{\en}{\end{equation}}
\newcommand{\bp}{\mathbf p}

\newcommand{\bk}{\mathbf k}
\newcommand{\br}{\mathbf r}

\newcommand \bel  {\begin{align}}
\newcommand \enl  {\end{align}}

\newcommand{\veps}{\varepsilon}
\newcommand{\eps}{\epsilon}

\newcommand{\up}{\uparrow}
\newcommand{\dn}{\downarrow}
\newcommand{\dg}{^\dagger}

\begin{document}

\title{Collisionless dynamics of the pairing amplitude in disordered superconductors}

\author{Maxim Dzero}
\address{Department of Physics, Kent State University, Kent, OH 44242, USA}

\begin{abstract}
I consider relaxation of the pairing amplitude in a disordered Bardeen-Cooper-Schrieffer (BCS) superconductor in the absence of the two-particle collisions. 
My main assumption is that nonmagnetic and magnetic disorder scattering rates are much smaller than the value of the superconducting pairing gap $\Delta_0$. I derive a system of nonlinear equations which describe the collisionless relaxation of the pairing amplitude following a quench of the pairing strength. I find that in a superconductor in which scattering on paramagnetic impurities is dominant, the pairing amplitude in a steady state varies periodically with time even for small deviations from equilibrium. It is shown that such a steady state emerges due to scattering on paramagnetic impurities which leads to a decrease in the value of the resonant frequency of the amplitude mode below $2\Delta_0$. 
\end{abstract}

\pacs{67.85.De, 34.90.+q, 74.40.Gh}

\date{\today}

\maketitle

\paragraph{Introduction.} Almost five decades ago, Volkov and Kogan published a theory of collisionless relaxation of the pairing gap in $s$-wave superconductors \cite{VolkovKogan1973}. The latter refers to a regime when the relevant time scale for the dynamics far exceeds order parameter relaxation time $\tau_\Delta=\hbar/\Delta_0$, but is much smaller than the relaxation time due to electron-electron collisions $\tau_{\textrm{ee}}\approx{\hbar\varepsilon_F/\Delta_0^2}$ ($\varepsilon_F$ is the Fermi energy). They have considered a model in which the Cooper pairing between the conduction electrons is mediated by their interaction with the acoustic phonons. By focusing on the time scales much longer than inverse value of the energy gap $\Delta_0$ in equilibrium as well as Debye frequency $\omega_D$, they have derived a system of equations describing the relaxation of an energy gap in clean superconductors. 
In the linear regime when the deviations from the equilibrium are almost negligible, the time dependence of the pairing amplitude had been found analytically, while the dynamics of the pairing amplitude for stronger deviations from equilibrium remained unknown \cite{VolkovKogan1973}.

A significant progress in the understanding of the collisionless pairing dynamics was made only thirty years after the work of Volkov and Kogan. The interest to this problem has been revived in the context of the superfluidity in the atomic condensates \cite{Spivak2004}. Indeed, in these systems one can induce collisionless dynamics by a sudden change of the magnetic field controlling the optical trap, which inevitably leads to a change of the pairing strength \cite{Ketterle2005}. Soon after that it was realized that the problem of finding the relaxation of the pairing amplitude for an arbitrary deviations from equilibrium admits an exact solution \cite{Enolski2005,Enolski2005a,Altshuler2005,Yuzbashyan2008} (see also Ref. \cite{qReview2015} for a comprehensive review). In particular, for strong enough deviations from equilibrium, it was found that the time dependence of the pairing amplitude at long times does not asymptote to a constant value, but remains periodic in time \cite{Spivak2004,Yuzbashyan2008}.

These theoretical developments have triggered the emergence of experiments which aimed to observe the evolution of the energy gap in superconducting films subject to an external electromagnetic pulses in terahertz frequency range \cite{Shimano2012,Demsar2013,Sherman2015}. Theoretical analysis of these experiments, however, typically relies on the results of the early theoretical works \cite{VolkovKogan1973,qReview2015,Foster2017}. Importantly, in the context of the methodology used in \cite{VolkovKogan1973} one usually completely neglects the effects of disorder, which is inevitably present in the superconducting samples and may affect the resulting nature of the steady state when the corresponding time scale due to disorder scattering $\tau_{\textrm{dis}}$ satisfies the condition $\tau_{\Delta}\ll\tau_{\textrm{dis}}\ll \tau_{\textrm{ee}}$. It is worth mentioning, that the effects of potential disorder has been recently studied in the context of the pump-probe setup \cite{Lorenzana1,Benfatto1}, however the effects of weak magnetic impurities on the dynamics of the amplitude Higgs mode have never been discussed so far. 
\begin{figure}[t]
\centering
\includegraphics[width=0.895\linewidth]{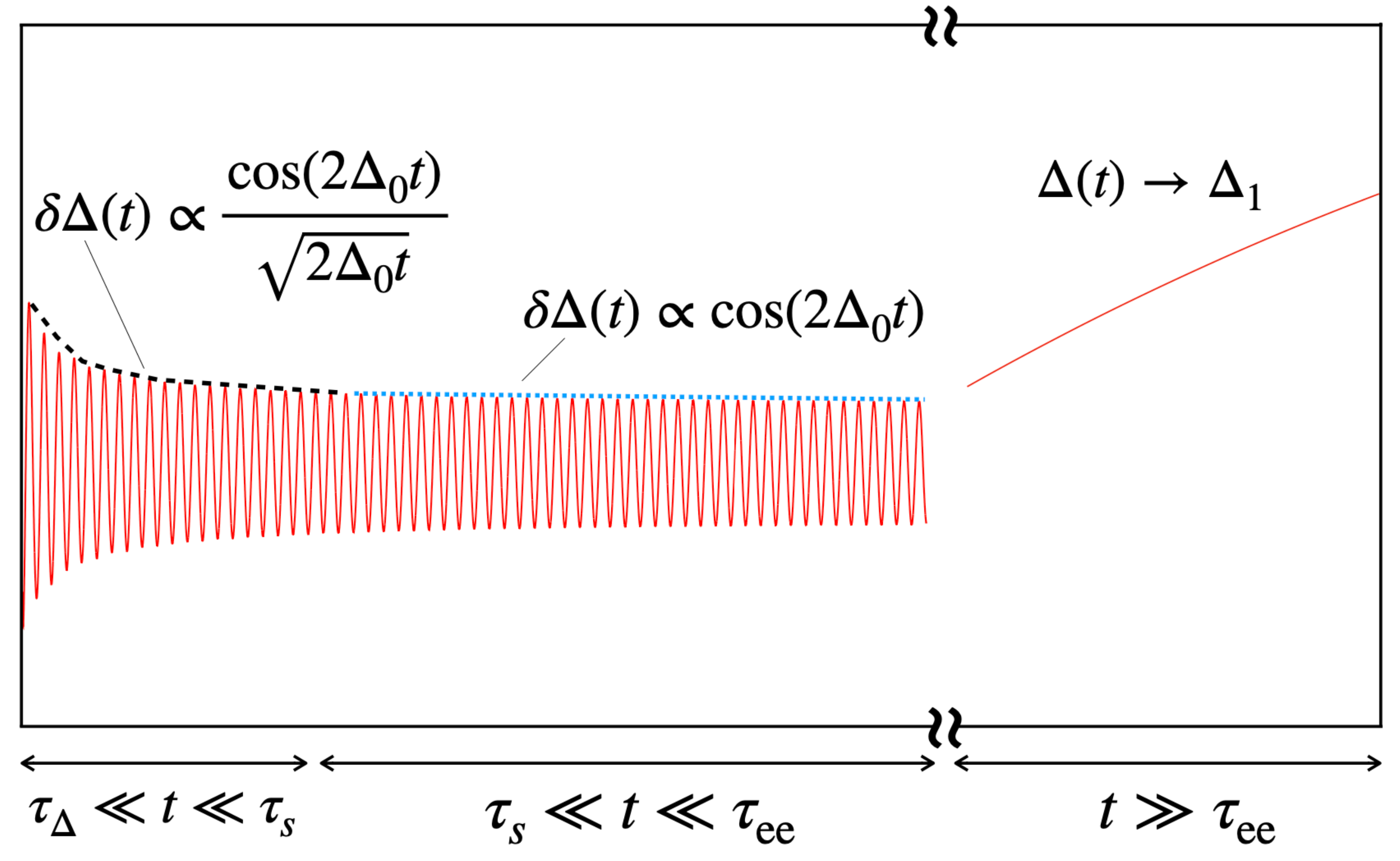}
\caption{Time evolution of the pairing amplitude in a conventional superconductor contaminated with a small amount of weak magnetic impurities following an abrupt (but small) change of the interaction strength $|\delta\lambda|/\lambda\ll 1$. The scattering on magnetic impurities is described by a relaxation time $\tau_s$. On a timescale 
$\tau_\Delta\ll t\ll\tau_s$ the order parameter $\delta\Delta(t)=\Delta(t)-\Delta_0$ oscillates with an amplitude which decays as $1/\sqrt{t}$. However, at longer times $\tau_s\ll t\ll\tau_{\textrm{ee}}$ the order parameter oscillated periodically with time. The amplitude of this oscillations is proportional to 
$1/(\tau_s\Delta_0)$. On a time scale $t\gg \tau_{\textrm{ee}}$ electron-electron scattering induces relaxation and the pairing amplitude reaches its new equilibrium value.}
\label{FigMain}
\end{figure}

In what follows I employ the Keldysh field-theoretical framework to derive a set of nonlinear equations for the dynamics of disordered superconductors in collisionless regime for the Bardeen-Cooper-Schrieffer (BCS) model of superconductivity \cite{BCS} including both nonmagnetic and paramagnetic disorder potentials. 
It is clear that dynamics can only be induced when the initial value of the pairing amplitude is different from the equilibrium one. One way to initiate the dynamics is to assume that the value of the  pairing strength has been instantaneously changed \cite{Spivak2004,Enolski2005,Enolski2005a}, so that by virtue of the self-consistency condition $\Delta(t=0)\not =\Delta_0$. Without loss of generality I will adopt this procedure here as well \cite{Foster2017}. Furthermore, from the exact solution of the Volkov-Kogan equations it is known that for small deviations from equilibrium  $\Delta(t)$ approaches a constant value \cite{qReview2015} at long times $\tau_\Delta\ll t\ll \tau_{\textrm{dis}}$. This behavior originates from the branch point at $\epsilon=2\Delta_0$ \cite{VolkovKogan1973,Spivak2004,qReview2015}. On this time scale the collision integrals, which we evaluate using the exact solution assuming that disorder is weak, should not and, in fact, they do not affect the dynamics.
At even longer times $t\sim\tau_{\textrm{dis}}\ll \tau_{\textrm{ee}}$, however, it is not \emph{a priori} clear whether the disorder scattering will produce changes to this steady state and it is precisely the question that I will address in this paper. 

In this Letter I demonstrate that in the presence of paramagnetic impurities the collisionless dynamics remains robust with respect to the dephasing processes, i.e. is dissipationless. Specifically, I find that already in the linear (Volkov-Kogan) regime and for a weak disorder, out-of-equilibrium dynamics of the pairing amplitude is described by a function which periodically oscillates with time. This is in stark contrast with the results of the earlier studies where much stronger deviations from equilibrium were required to find such type of a steady state \cite{qReview2015}. My result is schematically depicted in Fig. \ref{FigMain}.

\paragraph{Model and basic equations.}
We consider a model with the following Hamiltonian
\beg\label{Eq1}
\begin{split}
\hat{H}&=\sum\limits_{\alpha\beta}\int d^3\br\overline{\psi}_{\alpha}(\br)\left[{h}(-i{\vec \nabla})\delta_{\alpha\beta}+U_{\alpha\beta}(\br)\right]\psi_\beta(\br)\\&-g\int d^3\br\overline{\psi}_\up(\br)\overline{\psi}_\dn(\br){\psi}_\dn(\br){\psi}_\up(\br).
\end{split}
\en
Here $\psi_\sigma(\br)$ is an annihilation operator for a fermion with spin projection $\sigma=\pm1/2$, ${h}(-i{\vec \nabla})$ is a kinetic energy operator, $g$ is the coupling constant and
the last term accounts for disorder:
\beg\label{Disorder}
U_{\alpha\beta}(\br)=\sum\limits_ju(\br-\br_j)\delta_{\alpha\beta}+\left({\vec S}\cdot{\vec \sigma}_{\alpha\beta}\right)\sum\limits_l J(\br-\br_l).
\en
In (\ref{Disorder}) the summation is performed over the impurity sites and we assume that non-magnetic and paramagnetic impurities belong to different lattice sites and/or interstitials. 
The disorder potentials entering into this expression are described by the following correlators
$\langle u(\br)u(\br')\rangle_{\textrm{dis}}=\delta(\br-\br')/(2\pi\nu_F\tau_u)$, 
and $S(S+1)\langle J(\br)J(\br')\rangle_{\textrm{dis}}=\delta(\br-\br')/(2\pi\nu_F\tau_{s})$,
where $\nu_F$ is the single particle density of states at the Fermi level, $S$ is the spin of a paramagnetic impurity and the averaging is performed over disorder distribution. 

\paragraph{Equations of motion for the Green's functions.}
We consider the fermionic operators on Keldysh contour and introduce the correlation functions
$G_{\alpha\beta}^{(ab)}(1,2)=-i\langle\hat{T}_t\psi_{\alpha}(1_a)\overline{\psi}_\beta(2_b)\rangle$,
where $\psi_1=\psi_\up$, $\psi_2=\overline{\psi}_\dn$ and $a,b=1(2)$ refer to the top (bottom) parts of the Keldysh contour. As it directly follows from the definition of $G_{\alpha\beta}^{(ab)}(1,2)$, only three out of four functions (with respect to the Keldysh contour label) are independent: indeed, as it can be directly verified $\hat{G}^{(12)}+\hat{G}^{(21)}=\hat{G}^{(11)}+\hat{G}^{(22)}$. Hence we consider the retarded, advanced and the Keldysh propagators:
$\hat{G}^R(1,2)=\hat{G}^{(11)}(1,2)-\hat{G}^{(12)}(1,2)$, $\hat{G}^A(1,2)=\hat{G}^{(11)}(1,2)-\hat{G}^{(21)}(1,2)$ and $\hat{G}^K(1,2)=\hat{G}^{(11)}(1,2)+\hat{G}^{(22)}(1,2)$.
These functions satisfy the following relations $\left[G_{\alpha\beta}^{R(A)}\right]^*=-(-1)^{\alpha+\beta}G_{\overline{\alpha}\overline{\beta}}^{R(A)}$, 
$\left[G_{\alpha\beta}^K\right]^*=(-1)^{\alpha+\beta}G_{\overline{\alpha}\overline{\beta}}^{K}$, which we will use in what follows \cite{VolkovKogan1973}.

\paragraph{Dyson equations.} The equations of motion for the functions $G_{\alpha\beta}^{(ab)}(1,2)$ can be derived from the equations of motion for the fermionic operators. As a result one finds that these functions satisfy the Dyson equations:
\beg\label{DysonEqs}
\left[\hat{G}_0-\hat{\Sigma}\right]\circ\hat{G}=\hat{1}, \quad \hat{G}\circ\left[\hat{G}_0-\hat{\Sigma}\right]=\hat{1},
\en
where $\hat{G}_0$ denotes the bare Green's functions for a clean superconductor in the mean-field approximation.
\paragraph{Self-energy parts.} Self-energy parts $\hat{\Sigma}$ in Eqs. (\ref{DysonEqs}) can be obtained by perturbation theory \cite{Kamenev}. After performing averaging over disorder and using the correlators for the disorder potential (\ref{Disorder}) we found
\beg\label{Sigma}
\begin{split}
&{\Sigma}_{\sigma_1\sigma_2}^{(ij)}(1,2)=\frac{\delta(\br_1-\br_2)}{2\pi\nu_F\tau_s}\left(\hat{\gamma}_{im}^z{G}_{\sigma_1\sigma_2}^{(mn)}(1,2)\hat{\gamma}_{nj}^z\right)\\&+\frac{\delta(\br_1-\br_2)}{2\pi\nu_F\tau_u}\hat{\sigma}_{\sigma_1\sigma_3}^z
\left(\hat{\gamma}_{im}^z{G}_{\sigma_3\sigma_4}^{(mn)}(1,2)\hat{\gamma}_{nj}^z\right)\hat{\sigma}_{\sigma_4\sigma_2}^z,
\end{split}
\en
where the argument of the Green's function should be understood as $(1,2)=(x_1,x_2)$ with $x=(\br,t)$, $\hat{\sigma}^z$ and $\hat{\gamma}^z$ are Pauli matrices which act in Nambu and Keldysh contour spaces correspondingly (see Supplementary Materials). As it follows directly from the definition (\ref{Sigma}), these functions satisfy the relation $\hat{\Sigma}^{(11)}+\hat{\Sigma}^{(22)}+\hat{\Sigma}^{(12)}+\hat{\Sigma}^{(21)}=0$. Consequently, we introduce three independent self-energy functions 
$\hat{\Sigma}^R=\hat{\Sigma}^{(11)}+\hat{\Sigma}^{(12)}$,$\hat{\Sigma}^A=\hat{\Sigma}^{(11)}+\hat{\Sigma}^{(21)}$,
$\hat{\Sigma}^K=\hat{\Sigma}^{(12)}+\hat{\Sigma}^{(21)}$.
\paragraph{Equations of motion for the Keldysh function.} Having defined the self-energy part we are ready to write down the equation of motion for the Keldysh function. This is done in two steps (see Supplementary Materials for details on the derivation). First we obtain the equations of motion with respect to time $t=(t_1+t_2)/2$:
\beg\label{Eq4GKt}
\begin{split}
&[i\partial_{t}-{h}(1)+{h}^{*}(2)]G_{11}^{K}+{\Delta}(1)[G_{12}^{K}(1,2)]^*\\&+G_{12}^{K}(1,2)\overline{\Delta}(2)=I_{11}^K(1,2),\\
&[i\partial_{t}-{h}(1)-{h}^*(2)]G_{12}^{K}+G_{11}^{K}(1,2){\Delta}(2)\\&-{\Delta}(1)[G_{11}^{K}(1,2)]^*=I_{12}^K(1,2).
\end{split}
\en
Here we introduced the collision integrals
\beg\label{I11K}
\begin{split}
I_{\alpha\beta}^K(1,2)=\sum\limits_{\lambda}&\left(\Sigma_{\alpha\lambda}^{R}\circ G_{\lambda \beta}^{K}-\Sigma_{\alpha\lambda}^{K}\circ G_{\lambda \beta}^{A}\right.\\&\left.+G_{\alpha \lambda}^{R}\circ\Sigma_{\lambda \beta}^{K}-G_{\alpha \lambda}^{K}\circ\Sigma_{\lambda \beta}^{A}\right)(1,2).
\end{split}
\en
The second step consists in performing the Wigner transformation with respect to the relative time $\delta t=t_2-t_1$ and relative position $\delta\br=\br_2-\br_1$:
\beg\label{WT4G}
\check{G}(1,2)=\int\frac{d\veps}{2\pi}\int\frac{d^3\bp}{(2\pi)^3}\check{G}(t;\bp,\veps)e^{i\veps\cdot\delta t-i\bp\cdot\delta\br}.
\en
We use the same transformation for the collision integrals. 

Importantly, in order to compute the Wigner transform of the convolutions $C(1,2)=(A\circ B)(1,2)$ we use
$\hat{C}_{\bp\veps}(t)=\hat{A}_{\bp\veps}(t)e^{\frac{i}{2}(\stackrel{\leftarrow}{\partial}_\veps\stackrel{\rightarrow}{\partial}_t-\stackrel{\leftarrow}{\partial}_t\stackrel{\rightarrow}{\partial}_\veps)}\hat{B}_{\bp\veps}(t)\approx\hat{A}_{\bp\veps}(t)\hat{B}_{\bp\veps}(t)+(i/2)\left(\partial_\veps\hat{A}_{\bp\veps}\partial_t\hat{B}_{\bp\veps}
-\partial_t\hat{A}_{\bp\veps}\partial_\veps\hat{B}_{\bp\veps}\right)$. We note that only terms proportional to $G_{ab}^K$ in (\ref{I11K}) will have nonvanishing gradient contributions, since in a steady state that we consider the retarded and advanced propagators do not depend on $t$. Higher than linear derivatives of $\Sigma^{R(A)}$ with respect to $\veps$ will produce small pre-factors upon the integration over $\eps_\bk$ and for this reason their contributions can be ignored. In passing we note that the $t$-dependent contributions from $\Sigma_{\alpha\beta}^K$ produce an additional small pre-factor $\sim t^{-1/2}$ at long times and will also be ignored.  
\begin{figure}[t]
\centering
\includegraphics[width=0.95\linewidth]{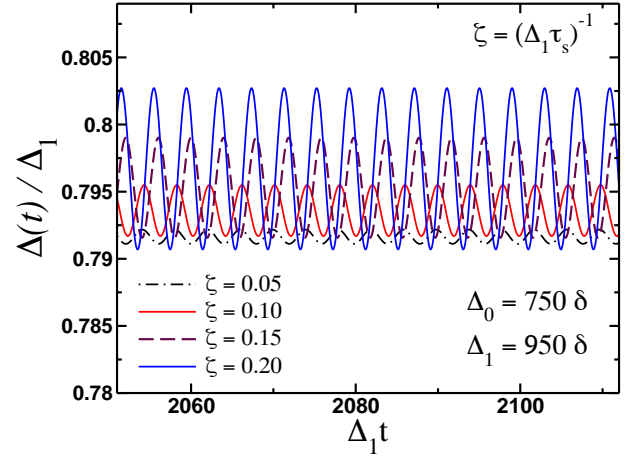}
\caption{Results of the numerical solution of the equations (\ref{MainEqs}) for the pairing amplitude $\Delta(t)$ at long times for various values of the dimenionless parameter $\zeta=1/\tau_s\Delta_1$ which quantifies the strength of paramagnetic disorder. These results have been obtained for the system of $N=15024$ equally spaced energy levels with the level spacing $\delta$. The dynamics was initiated by a sudden change of the dimensionless coupling constant from a value corresponding to the ground state with the pairing gap $\Delta_0$ to a new value corresponding to a ground state with the pairing gap $\Delta_1$.}
\label{Fig1}
\end{figure}

Given the form of the equations (\ref{Eq4GKt}), it will be convenient to work with the real functions ${\vec S}_\bp=(S_\bp^x,S_\bp^y,S_\bp^z)$ defined as follows
\beg\label{gpfp}
\begin{split}
&S_\bp^z(t)=i\int\limits_{-\infty}^\infty \frac{d\veps}{2\pi} G_{11}^K(\bp,\veps;t), \\
&S_\bp^x(t)+iS_\bp^y(t)=i\int\limits_{-\infty}^\infty \frac{d\veps}{2\pi} G_{12}^K(\bp,\veps;t).
\end{split}
\en
These quantities bear a clear analogy with Anderson pseudospins with the only exception that their norm $\vert{\vec S}_\bp\vert$ is not conserved by the evolution (see below).

We now use equations (\ref{Eq4GKt}) to derive the following equations for the components of ${\vec S}_\bp$:
\beg\label{MainEqs}
\begin{split}
&\frac{\partial S_\bp^x}{\partial t}-2\eps_\bp S_\bp^y=-\frac{\eps_\bp}{\tau\Delta_{\textrm{s}}}L_\bp^y(t), \\
&\frac{\partial S_\bp^y}{\partial t}+2\eps_\bp S_\bp^x+2\Delta(t)S_\bp^z=-\frac{\eps_\bp}{\tau_{\textrm{m}}E_\bp}\left({\cal W}_\bp-\cos\theta_\bp\right)
\\&+\frac{2L_{\bp}^z(t)}{\tau_s}+\frac{\eps_\bp}{\tau\Delta_{\textrm{s}}}
L_\bp^x(t), \\
&\frac{\partial S_\bp^z}{\partial t}-2\Delta(t)S_\bp^y=-\frac{2L_\bp^y(t)}{\tau_s},
\end{split}
\en
where $\eps_\bp\in[-\omega_D,\omega_D]$ are the single-particle energy levels, $\tau_{}^{-1}=\tau_s^{-1}+\tau_u^{-1}$, $\tau_{\textrm{m}}^{-1}=\tau_s^{-1}-\tau_u^{-1}$, ${\cal W}_\bp>0$ is a time-independent function which has a maximum at $\eps_\bp=0$ and decays to zero as $\eps_\bp\to\pm\infty$ (see Supplementary Materials). We have introduced the pairing function $\Delta(t)=-\Delta_{12}(1)$, which in its turn is determined self-consistently by
\beg\label{SelfConsfp}
\Delta(t)=\frac{\lambda}{2}\int\limits_{-\omega_D}^{\omega_D} S_\bp^x(t)d\eps_\bp.
\en
Here $\lambda=g\nu_F>0$ is the dimensionless coupling constant and $\omega_D$ is an ultraviolet cutoff, which reflects the retardation effects leading to the onset of superconductivity. In the derivation of the equations (\ref{MainEqs}) as well as in (\ref{SelfConsfp}) we have implicitly assumed that a superconductor is particle-hole symmetric, i.e. we consider a system with constant density of states $\nu(\eps)\approx\nu_F$. In Eqs. (\ref{MainEqs}) the components of ${\vec L}_\bp(t)$ describe the solution of the equations of motion in a clean superconductor $(\tau_{}\to\infty)$ at long times $\Delta(t\gg\tau_{\Delta})=\Delta_{\textrm{s}}$: 
\beg\label{PseudoSpins}
\begin{split}
{L}_\bp^x(t)&=\frac{\Delta_\textrm{s}}{E_\bp}\cos\theta_\bp+\frac{\eps_\bp}{E_\bp}\sin\theta_\bp\cos(2E_\bp t), \\
{L}_\bp^y(t)&=-\sin\theta_\bp\sin(2E_\bp t), \\
{L}_\bp^z(t)&=-\frac{\eps_\bp}{E_\bp}\cos\theta_\bp+\frac{\Delta_\textrm{s}}{E_\bp}\sin\theta_\bp\cos(2E_\bp t),
\end{split}
\en
where $E_\bp=(\eps_\bp^2+\Delta_{\textrm{s}}^2)^{1/2}$ and the components of vector ${\vec L}_\bp$ satisfy the normalization condition  ${\vec L}_\bp^2=1$. The expressions (\ref{PseudoSpins}) follow directly from (\ref{gpfp}) if instead of $G_{\alpha\beta}^K$ we use the expressions for the Keldysh propagators in the steady state (see Supplementary Materials). Given the perturbative nature of the calculation which lead to equations (\ref{MainEqs}) I would like to emphasize that the value of the pairing amplitude in the steady state $\Delta_{\textrm{s}}$ is taken to be equal to the one for a clean superconductor. We note that the equations of motion (\ref{MainEqs}) do not preserve the norm of ${\vec S}_\bp$ due to the pair breaking processes induced by the paramagnetic impurities. Finally, the definition of the function $\sin\theta_\bp$ can be found in the Supplementary Materials. 
\paragraph{Dynamics at long times.} We are now ready to investigate the collisionless dynamics following the quench of the pairing strength in a 
disordered superconductor. We start with the case when only paramagnetic impurities are present in a system, $\tau_u\to\infty$. 
In equilibrium $S_\bp^y=0$ while $S_{\bp}^{x}$ and $S_\bp^z$ can be determined from the equations of motion for $G^{R(A)}$ and are given by 
\beg\label{SpxInt}
\begin{split}
S_\bp^x(0)&=\frac{2}{\pi}\int\limits_0^\infty\frac{\left[R_u-\zeta_s\right][R_u^3-2\zeta_s]du}{R_u^4[(R_u-\zeta_s)^2+(\eps_\bp/\Delta_0)^2]},  \\
S_\bp^z(0)&=-\frac{2\eps_\bp}{\pi\Delta_0}\int\limits_0^\infty\frac{[R_u^3-2\zeta_s]du}{R_u^3[(R_u-\zeta_s)^2+(\eps_\bp/\Delta_0)^2]},
\end{split}
\en
where we introduced function $R_u=\sqrt{1+u^2}$ and dimensionless rate $\zeta_s=1/2\tau_s\Delta_0$ for brevity.
Furthermore, using equations (\ref{MainEqs}) it can be directly shown that in equilibrium the pseudospin components must satisfy $\eps_\bp S_\bp^x+\Delta_0 S_\bp^z=-w_\bp\eps_\bp/\tau_{s}(\eps_\bp^2+\Delta_0^2)^{1/2}$, where $w_\bp$ is a known function of $\eps_\bp$ and $\Delta_0$ is the energy gap in equilibrium computed using the self-consistent Born approximation (see Supplementary Materials for details).  

\begin{figure}[t]
\centering
\includegraphics[width=0.95\linewidth]{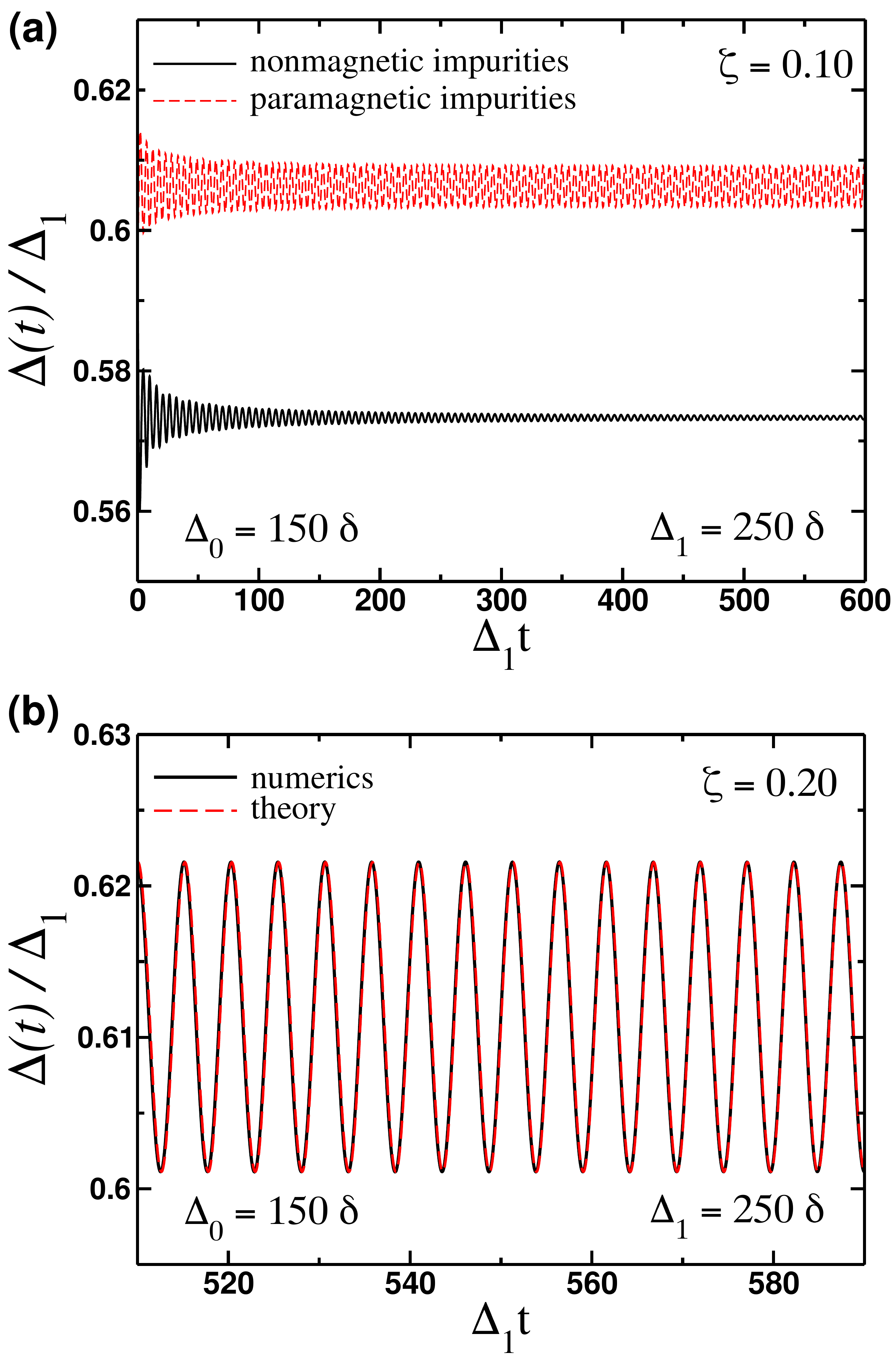}
\caption{Panel (a): Solution of the equations of motion for the two separate cases of (i) purely paramagnetic ($\tau_u\to\infty$) and (ii) purely nonmagnetic ($\tau_s\to\infty$) disorder. In the former case $\zeta=1/\tau_s\Delta_1$, the pairing amplitude periodically oscillates with time, while in the latter case $\zeta=1/\tau_u\Delta_1$ the amplitude of the oscillations decays as $t^{-1/2}$. Panel (b): Comparison between the results of the numerical solution for $\Delta(t)$ (paramagnetic disorder only) and the phenomenological expression $\Delta(t)=\Delta_0-\zeta[A+B\cos(2\Delta_0t+\varphi)]$. We used the following values of the parameters: $A\approx0.39$, $B\approx0.01$ and $\varphi\approx\pi/4$. We would like remind the reader that  these results hold on time scales shorter than $\tau_{\textrm{ee}}$.}
\label{Fig2}
\end{figure}

The results of the numerical solution of the equations (\ref{MainEqs}) following a small quenches of the pairing strength are shown in Fig. \ref{Fig1}. We immediately observe that the steady state with the oscillating $\Delta(t)$ emerges at times $t\sim\tau_s$. We also notice that the amplitude of the oscillations is proportional to $\zeta$. These results appear to be quite generic with respect to the magnitude and sign of the quench (i.e. when the pairing strength is slightly decreased) as well as initial conditions. We note that similar result has been recently reported, where the nondissipative Higgs mode appears due to the presence of long-range interactions in a superconductor coupled to a strongly driven cavity \cite{Cavity-Higgs}. Note also, that the amplitude of the oscillations is parametrically bigger for smaller values of the pairing gap in equilibrium, Fig. \ref{Fig2}. The full calculation of the steady state diagram for quenches of an arbitrary strength as well as strong disorder we leave for the future studies.
\paragraph{Discussion.} The appearance of the periodically oscillating solution can be understood as follows. Qualitatively, one may interpret this result an a way similar to the interpretation given in Ref. \cite{Levitov2007}: scattering induced by the paramagnetic impurities pushes the frequency of the amplitude mode inside the energy gap, so that the dephasing is fully suppressed and the mode becomes undamped. Indeed, simple calculation shows that for the frequency of the Higgs mode in this case ($\tau_u\to\infty$) we find
\beg\label{HiggsHiggsHurray}
\frac{\omega_{\textrm{Higgs}}}{2\Delta_0}=\left[1-\left(\frac{1}{\tau_s\Delta_0}\right)^2\right]^{1/2}.
\en 
The time dependence of the pairing amplitude will now be given by 
\beg\label{NewDLTt}
\delta\Delta(t)\approx \zeta e^{i\omega_{\textrm{Higgs}}t}+\int\limits_{-\infty}^\infty\frac{d\omega}{2\pi}A(\omega)e^{i\omega t}.
\en
The last term in this expression decays as $t^{-1/2}$ at long times and so only the first terms contributes. 

Eq. (\ref{HiggsHiggsHurray}) shows that contrary to our expectations in a superconductor with a pairing amplitude $\Delta_0$, it takes less than $2\Delta_0$ amount of energy to excite a Cooper pair. Let us also recall that in a superconductor contaminated with paramagnetic impurities, there is another energy scale $\Delta_{\textrm{th}}$ which represents the threshold for the single-particle excitations, i.e. the energy when the single particle density of states becomes nonzero for the first time \cite{AG1961} and I found that $\omega_{\textrm{Higgs}}>2\Delta_{\textrm{th}}$ (see Supplementary Materials). Therefore, equation (\ref{HiggsHiggsHurray}) introduces a completely new energy scale which describes the softening of the 'mass' of the Higgs mode. It seems that the physical processes which lead to the appearance of this energy scale are the same as the ones responsible for the appearance of $\Delta_{\textrm{th}}$.

In contrast, for a system in which only nonmagnetic disorder is present ($\tau_s\to\infty$), for the frequency of the Higgs mode we found
\beg\label{Boring}
\frac{\omega_{\textrm{Higgs}}}{2\Delta_0}=\left[1+\left(\frac{1}{\tau_u\Delta_0}\right)^2\right]^{1/2},
\en
so the frequency lies above the energy gap. This means that in this case the Higgs mode will become dissipative due to dephasing processes and, as a result, pairing amplitude asymptotes to a constant, $\tilde{\Delta}_{\textrm{s}}$, at long times. 

For the case when both nonmagnetic and magnetic disorder is present we find that the frequency of the amplitude mode falls below $2\Delta_0$ when $\tau_{\textrm{m}}>0$ and is above $2\Delta_0$ when $\tau_{\textrm{m}}<0$. This latter conclusion, which is based on the perturbative calculation, implies that the Higgs mode will become overdamped in either ballistic or dirty limits when $\tau_u\ll \tau_s$. As it turns out, in general this is not the case and it can be shown that even in diffusive superconductors the frequency of the resonant Higgs mode is less than 2$\Delta_0$. 

Comparison between these two behaviors, which are governed by Eqs. (\ref{HiggsHiggsHurray}) and (\ref{Boring}), is illustrated in Fig. \ref{Fig2}. In passing we note, that the Yu-Shiba-Rusinov bound states should not affect our results for the dynamics since the wave-functions describing these states remain orthogonal to the wave-functions corresponding to the continuous spectrum of the single-particle excitations. The same argument also applies to the existence of the tail states in the single particle density of states which are induced by the scattering on nonmagnetic impurities \cite{Meyer}. However, these effects may produce a shift in the value of $\omega_{\textrm{Higgs}}$ as well as the broadening if the Higgs resonance \cite{Armitage}. The detailed study of these effects as well as the question of whether these states affect the dynamics on the level of the collision integrals will be addressed separately. 

Finally, I note that apart from the fact that the physics I just described can be observed at time scales $t\ll\tau_{\textrm{ee}}$, Fig. \ref{FigMain}, it is also important to keep in mind that the steady state with the periodically oscillating pairing amplitude is intrinsically unstable towards developing spacial inhomogeneities \cite{Dzero2008} when the characteristic size of a sample $L$ is much larger than the coherence length $\xi=v_F/\Delta_0$. 
Thus, for the pairing amplitude to remain spatially inhomogeneous in a steady state with periodic oscillations, the value of the superconducting order parameter in equilibrium should be sufficiently small, so that the condition $\xi\gtrsim L$ is fulfilled. 

\paragraph{Conclusions.} In this work I have considered a problem of collisionless relaxation in a superconductor contaminated with nonmagnetic and paramagnetic impurities. I found that in the case when scattering on paramagnetic impurities is dominant, for even small deviations from equilibrium the corresponding steady state is described by the periodically oscillating pairing amplitude, which is an unmbiguous manifestation of the amplitude (Higgs) mode in a superconductor. 

It is well known that in a clean superconductor with the order parameter $\Delta_{00}$, it costs $\Delta_{00}$ amount of energy to excite a single particle and it costs $2\Delta_{00}$ amount of energy to excite a Cooper pair. Furthermore, in a disordered superconductor with the order parameter $\Delta_0$ in the presence paramagnetic impurities, in ballistic a regime it costs $\Delta_{\textrm{th}}<\Delta$ to create a single particle excitation \cite{AG1961}. The second main result of this Letter is that in this case it also costs $\omega_{\textrm{Higgs}}<2\Delta$ amount of energy to excite a Cooper pair. 

My results provide an avenue for detecting the Higgs mode in $s$-wave superconductors. 
Specifically, recent experimental studies of a superconductor NbN have convincingly demonstrated that in the pump-probe experimental setup, the intensity of the terahertz signal is peaked at $\omega_{\textrm{peak}}=2\Delta_0$ \cite{Armitage}. Based on the results of this work, I predict that by introducing small to moderate amount of paramagnetic impurities into NbN film and subjecting it to the electromagnetic pulse in the terahertz range of frequencies, one should observe the shift of the peak in the intensity of the signal from the expected value of $2\Delta_0$ to a smaller value $\omega_{\textrm{peak}}<2\Delta_0$. If several samples which differ by the amount of magnetic impurities, one should observe the decrease in the ratio $\omega_{\textrm{peak}}/2\Delta_0$ with an increase in the impurity concentrations. By combining this result with the results of the measurements which directly probe the single-particle density of states, one then can compare the value of $\omega_{\textrm{peak}}$ to $2\Delta_{\textrm{th}}$ and verify that $\omega_{\textrm{peak}}>2\Delta_{\textrm{th}}$. In case when there is an agreement between the experimental results and my theoretical predictions, it opens up a possibility for direct observation of the dynamics of the amplitude Higgs mode since, as it has been demonstrated in this work, the shift in the frequency of the Higgs mode means that its dynamics becomes undamped on a time scale $t\ll \tau_{\textrm{ee}}$. However, I am aware of the fact that the requirements for observing the pairing amplitude dynamics  are much more stringent due to the difficulties associated with controlling the value of the ratio of the relaxation times $\tau_s/\tau_{\textrm{ee}}$ and $\tau_u/\tau_{\textrm{ee}}$ experimentally.

\paragraph*{Acknowledgments.}
I would like to thank B. L. Altshuler, A. Balatsky, Y. Barlos, M. S. Foster,  M. Gershenson, A. Levchenko, D. Pesin, J. Schmalian and B. Z. Spivak for useful discussions. I am especially grateful to Emil Yuzbashyan for numerous illuminating discussions and insightful criticism of the results presented here. This work was financially supported by the National Science Foundation grant NSF-DMR-2002795.


\newpage
\begin{appendix}
\begin{center}
{\large\bf Supplementary Materials}
\end{center}
\vspace{0.1cm}

Here I provide additional technical details on the derivation of the main equations listed in the main text

\section{Expressions for the Green's functions in a steady state with $\Delta(t\to\infty)=\Delta_{\textrm{s}}$}
In this Section we derive the expressions for the retarded, advanced and Keldysh correlation functions in a steady state when the pairing amplitude asymptotes to a constant. Without loss of generality, we write the steady-state wave function in the following form 
\beg\label{UV}
\begin{split}
\left[\begin{matrix} u_\bp(t) \\ v_\bp(t) \end{matrix}\right]=s_\bp\left(\begin{matrix} U_\bp \\ V_\bp \end{matrix}\right)e^{iE_\bp t}+
c_\bp\left(\begin{matrix} V_\bp^* \\ -U_\bp^* \end{matrix}\right)e^{-iE_\bp t}.
\end{split}
\en
Here $c_\bp=\cos(\theta_\bp/2)$, $s_\bp=\sin(\theta_\bp/2)$, $E_\bp=\sqrt{\eps_\bp^2+\Delta_s^2}$, $\eps_\bp$ is the single-particle energy, $\Delta_{\textrm{s}}$ is the value of the pairing amplitude in the steady-state and
\beg\label{UpVp}
|U_\bp|^2=\frac{1}{2}\left(1+\frac{\eps_\bp}{E_\bp}\right), \quad |V_\bp|^2=\frac{1}{2}\left(1-\frac{\eps_\bp}{E_\bp}\right).
\en
Function $\sin(\theta_\bp/2)$ is known from the exact solution \cite{qReview2015a}. The physical meaning of this quantity is best understood using the framework of Anderson pseudospins, in which $\theta_\bp$ measures the deviation of a pseudospin from its equilibrium position, i.e. in equilibrium $\theta_\bp=0$. Averaging the correlation functions, which have been defined in the main text with the wave-function (\ref{UV}) yields:
\beg\label{SteadyStatePropsN}
\begin{split}
&G_{11}^{R}(\bp;t,t')=-i\theta(t-t')\left[\overline{u}_\bp(t)u_\bp(t')+\overline{v}_\bp(t')v_\bp(t)\right], \\
&G_{11}^{A}(\bp;t,t')=i\theta(t'-t)\left[\overline{u}_\bp(t)u_\bp(t')+\overline{v}_\bp(t')v_\bp(t)\right], \\
&G_{11}^{K}(\bp;t,t')=i\left[\overline{v}_\bp(t')v_\bp(t)-\overline{u}_\bp(t)u_\bp(t')\right].
\end{split}
\en
For the off-diagonal (anomalous) matrix elements we find
\beg\label{SteadyStatePropsA}
\begin{split}
&G_{12}^{R}(\bp;t,t')=i\theta(t-t')\left[\overline{u}_\bp(t)v_\bp(t')-\overline{u}_\bp(t')v_\bp(t)\right], \\
&G_{12}^{A}(\bp;t,t')=i\theta(t'-t)\left[\overline{u}_\bp(t')v_\bp(t)-\overline{u}_\bp(t)v_\bp(t')\right], \\
&G_{12}^{K}(\bp;t,t')=i\left[\overline{u}_\bp(t)v_\bp(t')+\overline{u}_\bp(t')v_\bp(t)\right].
\end{split}
\en
Expressions for the remaining functions can be computed with the help of the symmetry relations listed in the main text.

The Wigner transforms [Eq. (8) in the main text] of the Keldysh functions $G_{11}^{K}(\bp;t,t')$ and $G_{12}^{K}(\bp;t,t')$ can be easily obtained from the expressions above. For the normal (diagonal) components we find
\beg\label{WigGKFK}
\begin{split}
&{\cal G}_{\bp\veps}^K(t)=\int d(\delta t) G_{11}^{K}(\bp,\tau;t)e^{i\veps\cdot\delta t}=2\pi i(c_\bp^2-s_\bp^2)\\&\times\left(|U_\bp|^2\delta(\veps-E_\bp)-|V_\bp|^2\delta(\veps+E_\bp)\right)
\\&-4\pi ic_\bp s_\bp\left(\overline{U}_\bp \overline{V}_\bp e^{-2iE_\bp t}+U_\bp V_\bp e^{2iE_\bp t}\right)\delta(\veps)
\end{split}
\en
Similarly, for the anomalous component it obtains
\beg\label{WigFK}
\begin{split}
&{\cal F}_{\bp\veps}^K(t)=\int d(\delta t) G_{12}^{K}(\bp,\tau;t)e^{i\veps\cdot\delta t}\\&=2\pi i(s_\bp^2-c_\bp^2)\overline{U}_\bp V_\bp\left[\delta(\veps-E_\bp)+\delta(\veps+E_\bp)\right]\\&+4\pi i c_\bp s_\bp
\left[V_\bp^2e^{2iE_\bp t}-\overline{U}_\bp^2e^{-2iE_\bp t}\right]\delta(\veps).
\end{split}
\en
Lastly, we provide the expressions for the Wigner transforms of the functions $G_{11}^{R(A)}={\cal G}_{\bp\veps}^{R(A)}$ and $G_{12}^{R(A)}={\cal F}_{\bp\veps}^{R(A)}$:
\beg\label{WigGRA}
\begin{split}
&{\cal G}_{\bp\veps}^{R(A)}(t)=\frac{|U_\bp|^2}{\veps-E_\bp\pm i\delta}+\frac{|V_\bp|^2}{\veps+E_\bp\pm i\delta}, \\
&{\cal F}_{\bp\veps}^{R(A)}(t)=\frac{\overline{U}_\bp V_\bp}{\veps+E_\bp\pm i\delta}-\frac{\overline{U}_\bp V_\bp}{\veps-E_\bp\pm i\delta}.
\end{split}
\en
We use these expressions to compute the collision integrals in the Born approximation. For simplicity, we will assume that both $U_\bp$ and $V_\bp$ are real functions.

\section{Self-energy part} Disorder potentials induce the self-energy corrections to the single-particle correlation functions.
Here, for simplicity, I will consider the potential impurity and the Ising impurity 
\beg\label{Wdis}
W_{\alpha\beta}(\br)=\sum\limits_jw(\br-\br_j)+\left({\vec S}\cdot{\vec \sigma}_{\alpha\beta}\right)\sum\limits_iJ(\br-\br_i),
\en
where ${S}^z$ is the spin of magnetic impurity, $J(\br)$ are the magnetic scattering potentials and $w(\br)$ is the non-magnetic scattering potential. 
We assume that non-magnetic disorder potential has the correlation function
\beg\label{Ucorr}
\langle w(\br)w(\br')\rangle_{\textrm{dis}}=\frac{\delta(\br-\br')}{2\pi\nu_F\tau_u}.
\en
Magnetic disorder is described by the correlators
\beg\label{Scorr}
S(S+1)\langle J(\br)J(\br')\rangle_{\textrm{dis}}=\frac{\delta(\br-\br')}{6\pi\nu_F\tau_{s}}.
\en

\begin{figure}[t]
\centering
\includegraphics[width=0.95\linewidth]{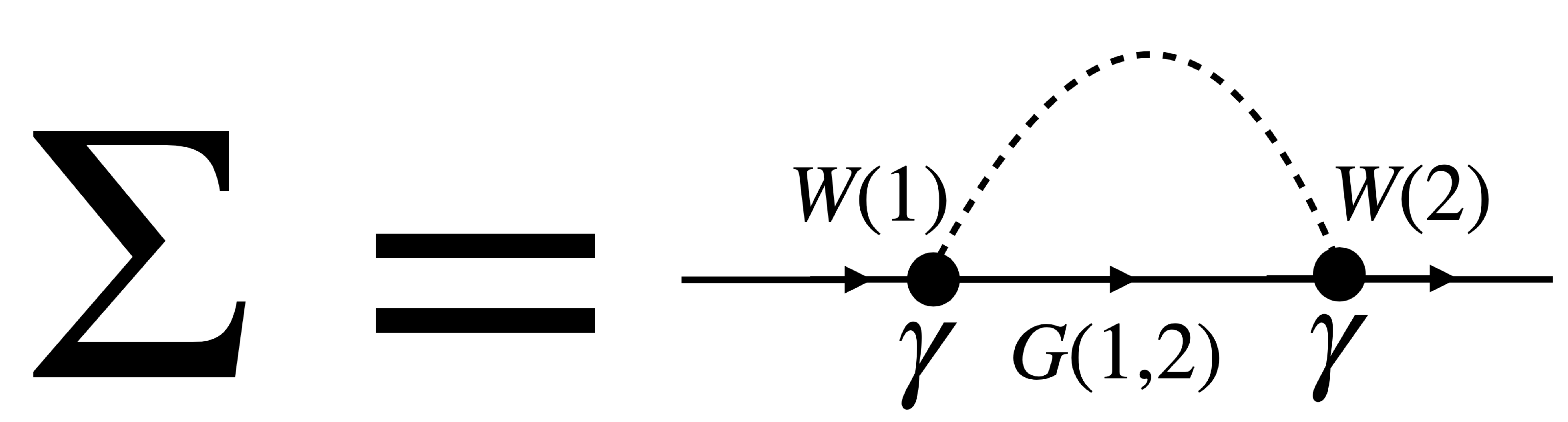}
\caption{Fermion self-energy diagram computed in the leading order with respect to nonmagnetic and paramagnetic impurities. }
\label{Supp-Fig1}
\end{figure}

Since we work in the Gor'kov-Nambu basis, in order to write down the expression for the self-energy, we first introduce
\beg\label{NambuSpin}
\hat{\Psi}=\left(\begin{matrix}\psi_{\up} \\ \overline{\psi}_\dn\end{matrix}\right).
\en
Next we write down the (no spin-flip) part of the action which describes disorder effects in the Nambu basis:
\beg\label{ActionDis}
\begin{split}
&S_{\textrm{dis}}[{\Psi}\dg,\Psi]\\&=\int\limits_{-\infty}^\infty dt\left\{\sum\limits_jw(\br-\br_j)\hat{\gamma}_{ab}^z
\left(\Psi_{\alpha a}\dg \hat{\rho}_{\alpha\beta}^{z} \Psi_{\beta b}\right)\right.\\&\left.+
S^z\sum\limits_jJ(\br-\br_i)\hat{\gamma}_{ab}^z
\left(\Psi_{\alpha a}\dg\Psi_{\alpha b}\right)
\right\}
\end{split}
\en
Here $\hat{\gamma}^z$ is the third Pauli matrix acting in the Keldysh basis and $\hat{\sigma}^z$ is the Pauli matrix in the Nambu basis. 
In order to include the effects of the spin-flip scattering, we expand the basis
\beg\label{Basis4}
\left(\begin{matrix} \psi_\up \\ \overline{\psi}_\dn \end{matrix}\right)\to 
\left(\begin{matrix} \psi_\up \\ \overline{\psi}_\dn \\ \psi_{\dn} \\ -\overline{\psi}_\up\end{matrix}\right), 
\left(\begin{matrix} \psi_{\dn} \\ -\overline{\psi}_\up\end{matrix}\right)=(i\hat{\sigma}_y\hat{\cal K})\left(\begin{matrix} \psi_\up \\ \overline{\psi}_\dn \end{matrix}\right)
\en
and $\hat{\cal K}$ is the complex-conjugation operator. 
Using the path-integral formalism it is then straightforward to expand the action in powers of the disorder potentials and use (\ref{Ucorr},\ref{Scorr}) to derive the following expression for self-energy
\beg\label{Self}
\begin{split}
&{\Sigma}_{\sigma_1\sigma_2}^{(ij)}(1,2)\\&=\frac{\delta(\br_1-\br_2)}{2\pi\nu_F\tau_u}\hat{\sigma}_{\sigma_1\sigma_3}^z
\left(\hat{\gamma}_{im}^z{G}_{\sigma_3\sigma_4}^{(mn)}(1,2)\hat{\gamma}_{nj}^z\right)\hat{\sigma}_{\sigma_4\sigma_2}^z\\&
+\frac{\delta(\br_1-\br_2)}{2\pi\nu_F\tau_s}\left(\hat{\gamma}_{im}^z{G}_{\sigma_1\sigma_2}^{(mn)}(1,2)\hat{\gamma}_{nj}^z\right).
\end{split}
\en
The expressions for the self-energy correspond to the diagrams in Fig. \ref{Supp-Fig1}.

\section{Dyson Equations} For the functions defined on the top side of the Keldysh contour we find
\beg\label{G11t1t2}
\begin{split}
&\left(i\frac{\partial}{\partial t_1}\delta_{\alpha\lambda}-\hat{h}_{\alpha\lambda}(1)\right)G_{\lambda\beta}^{(11)}-\hat{\Delta}_{\alpha\lambda}(1)G_{\lambda\beta}^{(11)}(1,2)\\&=\delta_{\alpha\beta}\delta(1-2)+\left(\Sigma_{\alpha\lambda}^{(1l)}\circ G_{\lambda\beta}^{(l1)}\right), \\
&\left(i\frac{\partial}{\partial t_2}\delta_{\lambda\beta}+\hat{h}_{\lambda\beta}^*(2)\right)G_{\alpha\lambda}^{(11)}+G_{\alpha\lambda}^{(11)}(1,2)\hat{\Delta}_{\lambda\beta}(2)\\&=-\delta_{\alpha\beta}\delta(1-2)-\left(G_{\alpha\lambda}^{(1l)}\circ\Sigma_{\lambda\beta}^{(l1)}\right)(1,2),
\end{split}
\en
where $\hat{\Sigma}$ are the self-energy parts due to disorder scattering, we used the mean-field approximation for the pairing term in the Hamiltonian [Eq. (1) in the main text] by introducing the pairing amplitude $\Delta(1)$, which must be determined self-consistently [see Eq. (10) in the main text] and 
matrices are given by 
\beg\label{Matrices}
\hat{h}=\left(\begin{matrix} h(1) & 0 \\ 0 & -h^*(1)\end{matrix}\right), \quad \hat{\Delta}(1)=\left(\begin{matrix} 0& \Delta(1) \\ \overline{\Delta}(1) & 0\end{matrix}\right).
\en
Expressions appearing in the right hand sides of Eqs. (\ref{G11t1t2}) imply the convolution with respect to both spacial and time coordinates as well as spin and Keldysh contour indices:
\beg\label{Conv}
\left(A\circ B\right)(1,1')=\int A(1,3)B(3,1')dx_3
\en
and $x_3=(\br_3,t_3)$. 
The Dyson equations for the functions defined on the bottom part of the Keldysh contour are
\beg\label{G22t1t2}
\begin{split}
&\left(i\frac{\partial}{\partial t_1}\delta_{\alpha\lambda}-\hat{h}_{\alpha\lambda}(1)\right)G_{\lambda\beta}^{(22)}-\hat{\Delta}_{\alpha\lambda}(1)G_{\lambda\beta}^{(22)}(1,2)\\&=-\delta_{\alpha\beta}\delta(1-2)-\left(\Sigma_{\alpha\lambda}^{(2l)}\circ G_{\lambda\beta}^{(l2)}\right)(1,2), \\
&\left(i\frac{\partial}{\partial t_2}\delta_{\lambda\beta}+\hat{h}_{\lambda\beta}^*(2)\right)G_{\alpha\lambda}^{(22)}+G_{\alpha\lambda}^{(22)}(1,2)\hat{\Delta}_{\lambda\beta}(2)\\&=\delta_{\alpha\beta}\delta(1-2)+\left(G_{\alpha\lambda}^{(2l)}\circ\Sigma_{\lambda\beta}^{(l2)}\right)(1,2).
\end{split}
\en
From these equations it is straightforward to derive the equations of motion for the Keldysh function $\hat{G}^K=\hat{G}^{(11)}+\hat{G}^{(22)}$. Next, we combine equations (\ref{G11t1t2}) and (\ref{G22t1t2}) to find the equations of motion with respect to $t_1$ and $t_2$ for $G_{\alpha\beta}^K$. Adding the resulting equations together and performing the Wigner transformation we arrive to the equations (5) in the main text. 

Lastly, we will also list the equations of motion for the functions $G_{\alpha\beta}^{(12)}$ which we will need in what follows as well:
\beg\label{G12RA}
\begin{split}
&\left(i\frac{\partial}{\partial t_1}\delta_{\alpha\lambda}-\hat{h}_{\alpha\lambda}(1)\right)G_{\lambda\beta}^{(12)}(1,2)-\hat{\Delta}_{\alpha\lambda}(1)G_{\lambda\beta}^{(12)}(1,2)\\&=\left(\Sigma_{\alpha\lambda}^{(1l)}\circ G_{\lambda\beta}^{(l2)}\right)(1,2), \\
&\left(i\frac{\partial}{\partial t_2}\delta_{\lambda\beta}+\hat{h}_{\lambda\beta}^*(2)\right)G_{\alpha\lambda}^{(12)}(1,2)+G_{\alpha\lambda}^{(12)}(1,2)\hat{\Delta}_{\lambda\beta}(2)\\&=\left(G_{\alpha\lambda}^{(1l)}\circ\Sigma_{\lambda\beta}^{(l2)}\right)(1,2).
\end{split}
\en
We recall the definition of the retarded function $\hat{G}^R=\hat{G}^{(11)}-\hat{G}^{(12)}$ to derive
\beg\label{GRt1t2}
\begin{split}
&\left(i\frac{\partial}{\partial t_1}\delta_{\alpha\lambda}-\hat{h}_{\alpha\lambda}(1)\right)G_{\lambda\beta}^{R}(1,2)-\hat{\Delta}_{\alpha\lambda}(1)G_{\lambda\beta}^{R}(1,2)\\&=\sum\limits_{\lambda}\left(\Sigma_{\alpha\lambda}^{R}\circ G_{\lambda\beta}^{R}\right)(1,2), \\
&\left(i\frac{\partial}{\partial t_2}\delta_{\lambda\beta}+\hat{h}_{\lambda\beta}^*(2)\right)G_{\alpha\lambda}^{R}(1,2)+G_{\alpha\lambda}^{R}(1,2)\hat{\Delta}_{\lambda\beta}(2)\\&=-\sum\limits_{\lambda}\left(G_{\alpha\lambda}^{R}\circ\Sigma_{\lambda\beta}^{R}\right)(1,2).
\end{split}
\en
Adding these equations together and performing the Wigner transformation yields the equations for the 'slow' dynamics of the retarded propagators:
\beg\label{dtGRa}
\begin{split}
&i\partial_tG_{11}^R(\bp\veps;t)+\Delta(t)\left[G_{12}^R(\bp\veps;t)-G_{21}^R(\bp\veps;t)\right]=0,\\
&\left(i\partial_t-2\eps_\bp\right)G_{12}^R(\bp\veps;t)\\&+\Delta(t)\left[G_{22}^R(\bp\veps;t)-G_{11}^R(\bp\veps;t)\right]=I_{12}^R(\veps).
\end{split}
\en
The expressions for the collision integrals can be computed exactly without involving the gradient expansion:
\beg\label{CollisionsIR}
\begin{split}
I_{12}^R(\veps)=\Sigma_{12}^R(\veps)\left[\tilde{\cal G}_{\bp\veps}^R-{\cal G}_{\bp\veps}^R\right],
\end{split}
\en
where 
\beg\label{tGR}
\tilde{\cal G}_{\bp\veps}^R=\frac{|U_\bp|^2}{\veps+E_\bp+i0}+\frac{|V_\bp|^2}{\veps-E_\bp+i0}.
\en
One can easily check that the steady state propagators (\ref{WigGRA}) satisfy these equations on a time scale $\tau_\Delta\ll t\ll \tau_{\textrm{dis}}$.
Since the collision integral $I_{11}^R=0$, the steady state propagators satisfy the first equation (\ref{dtGRa}).

\section{Anderson pseudospins in disordered superconductors}
In this Section we are going to derive the relation between the $x$- and $z$-components of the Anderson pseudospins. We will need a function \beg\label{Wp}
{\cal W}_\bp=\frac{\Delta_s}{\pi\nu_F}\int\frac{d^3\bk}{(2\pi)^3}\frac{\cos\theta_\bk}{2E_\bk}\left(\frac{1}{E_\bk+E_\bp}-{\cal P}\frac{1}{E_\bk-E_\bp}\right)
\en
and the second term under the integral should be evaluated as a principal value. In equilibrium $\cos\theta_\bk=1$, $\Delta_{\textrm{s}}=\Delta_0$ and we readily find
\beg\label{EqSmallDis}
w_\bp={\cal W}_\bp(\theta_\bp=0)=\frac{2\Delta_0}{\pi \eps_\bp}
\sinh^{-1}\left(\frac{\eps_\bp}{\Delta_0}\right).
\en
We also note that in equilibrium the pseudospins must satisfy
\beg\label{RelEqui}
\eps_\bp S_\bp^x+\Delta_0 S_\bp^z=\frac{w_\bp}{\tau_{\textrm{m}}}L_\bp^z.
\en
This relation follows directly from the second equation (9) in the main text.
Because the pseudospins do not satisfy normalization condition, in order to determine $S_\bp^{x}$ and $S_\bp^z$ in equilibrium, we need to use additional equations. 
\subsection{Equations for $G^{R(A)}$} To compute the expressions for $S_\bp^x$ and $S_\bp^z$ in equilibrium, we will use the equations of motion for $G_{ab}^R(\bp;t_1,t_2)$ with respect to the relative time $\delta t=t_1-t_2$. It is useful to keep in mind that in equilibrium the Keldysh component of $\hat{G}$ satisfies
\beg\label{My2PlayRel}
\hat{G}^K(\bp,\veps)=\left[1-2\vartheta(\veps)\right]\left[\hat{G}^R(\bp,\veps)-\hat{G}^A(\bp,\veps)\right],
\en
where ${\vartheta}(\eps)$ is the Heaviside step function. From this expression it is clear that if $\hat{G}^{R(A)}(\bp,\veps)$ are known, than we can compute $\hat{G}^K(\bp,\veps)$ and obtain the configuration of the Anderson pseudospins using equations (8) in the main text.

Let us adopt the following compact notations
\beg\label{ShortNots}
G=G_{11}^R, \quad \overline{G}=G_{22}^R, \quad F=G_{12}^R=G_{21}^R.
\en
Equations for the functions $G$ and $F$ are
\beg\label{MyMainREqs}
\begin{split}
&(\veps-\eps_\bp)G(\bp,\veps)+\Delta_0F(\bp,\veps)=1+J_{11}(\bp,\veps), \\
&2\veps F(\bp,\veps)+\Delta_0\left[G(\bp,\veps)+\overline{G}(\bp,\veps)\right]=J_{12}(\bp,\veps), \\
&(\veps+\eps_\bp)\overline{G}(\bp,\veps)+\Delta_0F(\bp,\veps)=1+J_{22}(\bp,\veps).
\end{split}
\en
Here $J_{ab}(\bp,\veps)$ are the collision integrals, which will be computed in what follows. In deriving these equations we took into account that in the presence of the particle-hole symmetry $G_{12}^R=G_{21}^R$.  
In a clean superconductor, the retarded propagators are given by 
\beg\label{RetardedClean}
\begin{split}
&{\cal G}_{11}^R(\bp,\veps)=\frac{|U_\bp|^2}{\veps-E_\bp+i0}+\frac{|V_\bp|^2}{\veps+E_\bp+i0}, \\
&{\cal G}_{22}^R(\bp,\veps)=\frac{|U_\bp|^2}{\veps+E_\bp+i0}+\frac{|V_\bp|^2}{\veps-E_\bp+i0}, \\
&{\cal G}_{12}^R(\bp,\veps)=\frac{\Delta_0}{2E_\bp}\left(\frac{1}{\veps+E_\bp+i0}-\frac{1}{\veps-E_\bp+i0}\right),
\end{split}
\en
where $E_\bp=\sqrt{\eps_\bp^2+\Delta_0^2}$.
It is straightforward to verify that functions (\ref{RetardedClean}) satisfy (\ref{MyMainREqs}) when all $J_{ab}(\bp,\veps)=0$. It is worth noting here that the expressions for the propagators in equilibrium are formally the same as the ones in the steady state. 
\subsection{Self-consistent Born approximation} We consider the expressions for the collision integral $J_{11}(\bp,\veps)$:
\beg\label{J11}
J_{11}(\bp,\veps)=\Sigma_{11}^R(\veps){\cal G}_{11}^R(\bp,\veps)+\Sigma_{12}^R(\veps){\cal G}_{12}^R(\bp,\veps).
\en
Let us now replace ${\cal G}_{ab}^R(\bp,\veps)$ with the exact functions. As we will show below this corresponds to the self-consistent Born approximation. The equations (\ref{MyMainREqs}) are:
\beg\label{FirstEq4GR}
\begin{split}
&(\tilde{\veps}-\eps_\bp)G(\bp,\veps)+\tilde{\Delta}_\veps\cdot F(\bp,\veps)=1, \\
&(\tilde{\veps}+\eps_\bp)\overline{G}(\bp,\veps)+\tilde{\Delta}_\veps\cdot F(\bp,\veps)=1, \\
&2\tilde{\veps}F(\bp,\veps)+\tilde{\Delta}_\veps\cdot\left[{G}(\bp,\veps)+\overline{G}(\bp,\veps)\right]=0,
\end{split}
\en
where 
\beg\label{tepstD}
\tilde{\veps}=\veps-\Sigma_{11}^R(\veps), \quad \tilde{\Delta}_\veps=\Delta_0-\Sigma_{12}^R(\veps).
\en
As a next step, we solve (\ref{FirstEq4GR}) for $G(\bp,\veps)$ and $F(\bp,\veps)$:
\beg\label{Solve4GF}
\begin{split}
G(\bp,\eps)&=\frac{\tilde{\veps}+\eps_\bp}{\tilde{\veps}^2-\eps_\bp^2-\tilde{\Delta}_\veps^2}, \\
F(\bp,\eps)&=-\frac{\tilde{\Delta}_\veps}{\tilde{\veps}^2-\eps_\bp^2-\tilde{\Delta}_\veps^2}.
\end{split}
\en

For the self-energies computed using (\ref{RetardedClean}) we have found
\beg\label{SigmasR}
\begin{split}
\Sigma_{11}^R(\veps)&=\left(\frac{i}{2\tau_{\textrm{dis}}}\right)\frac{(\veps+i0)}{\sqrt{(\veps+i0)^2-\Delta_0^2}}, \\
\Sigma_{12}^R(\veps)&=-\left(\frac{i}{2\tau_{\textrm{m}}}\right)\frac{\Delta_0}{\sqrt{(\veps+i0)^2-\Delta_0^2}}.
\end{split}
\en
If we now compute the self-energies using exact functions (\ref{Solve4GF}) we will find the same expressions as (\ref{SigmasR}) where $\tilde{\veps}$ replaces ${\veps}$ and $\Delta_0$ is replaced with $\tilde{\Delta}_\veps$. 
Going back to (\ref{tepstD}) yields the following self-consistency equations for the quantities $\tilde{\veps}$ and $\tilde{\Delta}_\veps$:
\beg\label{Replace}
\begin{split}
\tilde{\veps}=\veps+\frac{u_\veps}{2\tau\sqrt{1-u_\veps^2}}, ~\tilde{\Delta}_\veps=\Delta_0-\frac{1}{2\tau_{\textrm{m}}\sqrt{1-u_\veps^2}},
\end{split}
\en
where $u_\veps={\tilde{\veps}}/{\tilde{\Delta}_\veps}$.
From this expressions we can also derive the relation for the ratio:
\beg\label{epsdlt}
\frac{\veps}{\Delta_0}=u_\veps\left(1-\frac{1}{\tau_s\Delta_0\sqrt{1-u_\veps^2}}\right).
\en
\paragraph{Anderson theorem.} To compare these expressions with the Abrikosov-Gor'kov result \cite{AG1961}, we formally replace $\tilde{\veps}=i\tilde{\omega}_n$, $\tilde{\Delta}_\veps=\tilde{\Delta}_n$ and $\veps=i\omega_n$, where $\omega_n=\pi T(2n+1)$ is the fermionic Matsubara frequency. It follows
\beg\label{ReAGwn}
\begin{split}
&\tilde{\omega}_n=\omega_n+\frac{u_n}{2\tau\sqrt{1+u_n^2}}, \\ 
&\tilde{\Delta}_n=\Delta_0-\frac{1}{2\tau_{\textrm{m}}\sqrt{1+u_n^2}}.
\end{split}
\en
Comparing these equations with the equations (41) and (42) in Ref. \cite{AG1961} we have
\beg\label{tau1tau2}
\frac{1}{\tau_1}=\frac{1}{\tau}=\frac{1}{\tau_u}+\frac{1}{\tau_s}, \quad \frac{1}{\tau_2}=-\frac{1}{\tau_{\textrm{m}}}=\frac{1}{\tau_u}-\frac{1}{\tau_s}.
\en
Note that when $\tau_s\to\infty$ it follows from (\ref{ReAGwn}) that $u_n=(\tilde{\omega}_n/\tilde{\Delta}_n)=\omega_n/\Delta_0$. Indeed 
\beg\label{AndTheorem}
u_n\Delta_0+\frac{u_n}{2\tau_u\Delta_0\sqrt{1+u_n^2}}=\omega_n+\frac{u_n}{2\tau_u\sqrt{1+u_n^2}}, 
\en
so that
\beg\label{Part2}
u_n=\frac{\omega_n}{\Delta_0}.
\en
Let us consider the self-consistency condition (Eq. (10) in the main text):
\beg\label{SelfSupp}
\begin{split}
&\Delta_0=\frac{\lambda}{2} T\sum\limits_{\omega_n}\sum\limits_{\bp}\frac{\tilde{\Delta}_n}{\tilde{\omega}_n^2+\tilde{\Delta}_n^2+\eps_\bp^2}\\&=
\left(\frac{\pi\lambda}{2}\right)T\sum\limits_{\omega_n}\frac{\tilde{\Delta}_n}{\sqrt{\tilde{\omega}_n^2+\tilde{\Delta}_n^2}}\\&=
\left(\frac{\pi\lambda}{2}\right)T\sum\limits_{\omega_n}\frac{\Delta_0}{\sqrt{\omega_n^2+\Delta_0^2}}
\end{split}
\en
and on the last step we used (\ref{Part2}).
This result is a manifestation of Anderson theorem: 
small amount of non-magnetic impurities does not affect the magnitude of the  superconducting energy gap. 
\subsection{Anderson pseudospins}
To compute the configuration of the Anderson pseudospins we will use expression (\ref{My2PlayRel}). Specifically, we insert (\ref{My2PlayRel}) into Eqs. (8) in the main text and recall that the resulting integration over $\veps$ is equivalent to performing the summations over fermionic Matsubara frequencies by virtue of the following relation
\beg\label{ContourTrick}
\begin{split}
&T\sum\limits_{\omega_n}f(\omega_n)=\oint\frac{dz}{4\pi i}\tanh\left(\frac{z}{2T}\right) f(z)\\&=\int\limits_{-\infty}^\infty\frac{dz}{4\pi i}\tanh\left(\frac{z}{2T}\right)
\left(f(z+i0)-f(z-i0)\right).
\end{split}
\en

Thus, in the expressions (\ref{Replace}) above we formally replace $\veps$ with $i\omega_n=i\pi T(2n+1)$, $\tilde{\veps}$ with $\tilde{\omega}_n$ and $\tilde{\Delta}_\veps$ with $\tilde{\Delta}_n$ arriving to Eqs. (\ref{ReAGwn}).
Then, we consider the following Matsubara summations:
\beg\label{Spx}
\begin{split}
{\cal S}_\bp^x&=2T\sum\limits_{\omega_n}\frac{\tilde{\Delta}_n}{\tilde{\omega}_n^2+\tilde{\Delta}_n^2+\eps_\bp^2}, \\ 
{\cal S}_\bp^z&=-2T\sum\limits_{\omega_n}\frac{\eps_\bp}{\tilde{\omega}_n^2+\tilde{\Delta}_n^2+\eps_\bp^2}.
\end{split}
\en 
In the case of $T=0$ we can just replace the summation with an integral over the frequencies according to
\beg\label{Sum2Int}
T\sum\limits_{\omega_n}\to\int\frac{d\Omega}{2\pi}.
\en
By combining two equations (\ref{ReAGwn}) we can express the ratio $\omega_n/\Delta_0$ in terms of $u_n$:
\beg\label{MainExpress}
\frac{\omega_n}{\Delta_0}=u_n-\frac{u_n}{\tau_s\Delta_0\sqrt{1+u_n^2}}.
\en

It will be convenient to change the integration over $d\Omega$ to the one over $du$:
\beg\label{dOmegadu}
\frac{d\Omega}{\Delta_0}=du\left(1-\frac{1}{\tau_s\Delta_0}\frac{1}{(1+u^2)^{3/2}}\right).
\en

For the function ${\cal S}_\bp^x$ we have the following expression
\beg\label{SpxInt}
{\cal S}_\bp^x=\frac{2}{\pi}\int\limits_0^\infty\frac{du}{R^4(u)}\frac{\left[R(u)-\zeta_{\textrm{m}}\right][R^3(u)-\zeta_s]}{[(R(u)-\zeta_{\textrm{m}})^2+(\eps_\bp/\Delta)^2]}, 
\en
where $R(u)=\sqrt{1+u^2}$ and we introduced the dimensionless parameters
\beg\label{ParamsParams}
\zeta_s=\frac{1}{\tau_s\Delta_0}, \quad \zeta_{\textrm{m}}=\frac{1}{2\tau_{\textrm{m}}\Delta_0}.
\en
It is easy to verify that in the clean limit ${\cal S}_\bp^x=\Delta_{0}/E_\bp$. If we insert this expression into the self-consistency condition [Eq. (10) in the main text],
we readily find
\beg\label{SimpleCalc}
\log\left(\frac{\Delta_0}{\Delta_0^{(0)}}\right)=-\frac{\pi}{4\tau_s\Delta_0},
\en
where $\Delta_0^{(0)}$ is the energy gap in a clean superconductor. 

Similarly, the expression for the function ${\cal S}_\bp^z$ is
\beg\label{SpzInt}
{\cal S}_\bp^z=-\frac{2\eps_\bp}{\pi\Delta_0}\int\limits_0^\infty\frac{du}{R^3(u)}\frac{[R^3(u)-\zeta_s]}{[(R(u)-\zeta_{\textrm{m}})^2+(\eps_\bp/\Delta_0)^2]}.
\en
Comparing these two expressions, it follows
\beg\label{RelationMain}
\begin{split}
{\cal S}_\bp^z&=-\frac{\eps_\bp}{\Delta_0}S_\bp^x-\frac{1}{\tau_{\textrm{m}}\Delta_0}\left(\frac{2\eps_\bp}{\pi\Delta_0}\right)\\&\times\int\limits_0^\infty\frac{du}{R^4(u)}\frac{[R^3(u)-\zeta_s]}{[(R(u)-\zeta_{\textrm{m}})^2+(\eps_\bp/\Delta_0)^2]}.
\end{split}
\en
In the limit of weak disorder we can completely neglect the relaxation rates under the integral. Then, the latter evaluates to 
\beg\label{EvalInt}
\frac{2}{\pi\Delta_0}\int\limits_0^\infty\frac{du}{\sqrt{1+u^2}[1+u^2+(\eps_\bp/\Delta)^2]}=\frac{{w}_\bp}{E_\bp}.
\en
Inserting this expression into (\ref{RelationMain}) yields
\beg\label{MainAG}
\eps_\bp {\cal S}_\bp^x+\Delta_0 {\cal S}_\bp^z=-\frac{\eps_\bp{w}_\bp}{\tau_{\textrm{m}}\sqrt{\eps_\bp^2+\Delta_0^2}}.
\en
This expression is, in fact, identical to (\ref{RelEqui}). To summarize, expressions (\ref{SpxInt}) and (\ref{SpzInt}) along with $S_\bp^y=0$ determine the configuration of pseudo-spins in a superconductor with paramagnetic impurities in the particle-hole symmetric case. 

\section{Frequency of the Higgs mode}
The sole purpose of this Section is to verify that the paramagnetic impurities push the frequency of the pairing amplitude mode (Higgs mode) below $2\Delta_0$. We remind the reader the $\Delta_0$ refers to the value of the pairing amplitude in equilibrium for disordered superconductor.  We first discuss the limit when only paramagnetic impurities are present, $\tau_u\to\infty$. 

In order to find the frequency of the Higgs mode we linearize the equations of motion [Eqs. (9) in the main text] with the ansatz $\Delta(t)=\Delta_0+e^{-i\omega t}\delta\Delta_\omega$ and ${\vec S}_\bp={\vec {\cal S}}_\bp+e^{-i\omega t}\delta{\vec S}_{\bp\omega}$. Expressing $\delta S_{\bp\omega}^x$ in terms of $\delta\Delta_\omega$ from the self-consistency equation we find
\beg\label{wHiggs1}
\delta\Delta_\omega=-\frac{\lambda}{2}\int\limits_{-\omega_D}^{\omega_D}\frac{\eps_\bp{\cal S}_{\bp}^zd\eps_\bp}{\eps_\bp^2+\Delta_0^2-(\omega/2)^2}\delta\Delta_\omega.
\en
Assuming $\delta\Delta_\omega\not=0$ and replacing $2/\lambda$ using the self-consistency condition transforms (\ref{wHiggs1}) into
\beg\label{wHiggs2}
\begin{split}
\int\limits_{-\omega_D}^{\omega_D}\left[{\cal S}_{\bp}^x+\frac{\eps_\bp\Delta_0{\cal S}_{\bp}^z}{\eps_\bp^2+\Delta_0^2-(\omega/2)^2}\right]d\eps_\bp=0.
\end{split}
\en
In the second term under the integral we replace $\Delta_0{\cal S}_\bp^z$ using (\ref{MainAG}) with the following result:
\beg\label{wHiggs3}
\begin{split}
&\int\limits_{-\infty}^{\infty}\frac{[\Delta_0^2-(\omega/2)^2]{\cal S}_{\bp}^xd\eps_\bp}{\eps_\bp^2+\Delta_0^2-(\omega/2)^2}
\\&=\frac{1}{\tau_s}\int\limits_{-\infty}^{\infty}\frac{\eps_\bp^2w_\bp d\eps_\bp}{E_\bp\left[\eps_\bp^2+\Delta_0^2-(\omega/2)^2\right]}.
\end{split}
\en
Here we set the integration limits to infinity, since the integrals are converging. In the right-hand side of this equation we use the integral representation 
of the function $w_\bp$, Eq. (\ref{EvalInt}), to integrate over $\eps_\bp$ first and then perform the remaining integration over $u$. As a result, equation (\ref{wHiggs3}) becomes
\beg\label{wHiggs4}
\begin{split}
&(\Delta_0a_\omega)^2\int\limits_{-\infty}^{\infty}\frac{{\cal S}_{\bp}^xd\eps_\bp}{\eps_\bp^2+\Delta_0^2-(\omega/2)^2}
=\frac{2\cos^{-1}(a_\omega)}{\tau_s\sqrt{1-a_\omega^2}},
\end{split}
\en
where $a_\omega=\sqrt{1-(\omega/2\Delta_0)^2}$. Clearly, in the clean limit we recover familiar result $\omega=2\Delta_0$. Since we are looking for the linear in $\tau_s^{-1}$ correction to $\omega$, it will suffice to replace ${\cal S}_\bp^x$ with its expression in the clean limit. Performing the remaining integration over $\eps_\bp$ yields 
\beg\label{AuxInt}
\int\limits_{-\infty}^{\infty}\frac{\Delta_0^2d\eps_\bp}{\sqrt{\eps_\bp^2+\Delta_0^2}(\eps_\bp^2+\Delta_0^2a_\omega^2)}=\frac{2\cos^{-1}(a_\omega)}{a_\omega\sqrt{1-a_\omega^2}}.
\en
Inserting this expression into (\ref{wHiggs4}) gives the following expression for the frequency of the Higgs mode:
\beg\label{SpxLin}
\frac{\omega}{2\Delta_0}=\sqrt{1-\left(\frac{1}{\tau_s\Delta_0}\right)^2},
\en
which is manifestly quadratic in $(\tau_s\Delta_0)^{-1}$ for $\tau_s\Delta_0\gg 1$.

It is important to keep in mind that in an $s$-wave superconductor with paramagnetic impurities, the gap in the energy spectrum for the single-particle excitations $\Delta_{\textrm{th}}$ is different from the pairing amplitude $\Delta_0$  and it is given by \cite{AG1961}
\beg\label{wth}
\Delta_{\textrm{th}}=\Delta_0\left[1-\left(\frac{1}{\tau_s\Delta_0}\right)^{2/3}\right]^{3/2}.
\en
Note that this result does not follow directly from the equilibrium expressions for the pseudospins. 
Moreover, our calculation demonstrates that the single-particle energy gap $\Delta_{\textrm{th}}$ does not appear and the Cooper pair excitation energy $\omega$ is
still determined by $\Delta_0$. In this sense it is analogous to the superfluid density, which is also determined by $\Delta_0$. Lastly, we expect that the single-particle states with energies above $\Delta_{\textrm{th}}$ will contribute to the dynamics at times $\tau_{\veps}\sim E_F/\Delta_{\textrm{th}}^2$ which is assumed to be much longer than $\tau_{s}$. 

\section{Steady state configuration of the pseudospins in the clean limit}
In the linear regime, i.e. when the magnitude of the quench is weak, at long times the system reaches the steady state. In that steady state each pseudospin precesses around its effective magnetic field ${\vec B}_\bp=2(\Delta_{\textrm{s}}^{(x)}, \Delta_{\textrm{s}}^{(y)}, -\eps_\bp)$. 
In the BCS model, we have the following expression for the
angle between the effective magnetic field at $t\to\infty$ and the pseudospin at the single particle energy level 
$\eps_{\mathbf p}$:
\beg\label{SinTheta}
\sin^2\theta_\bp=\frac{1}{2\pi^2\rho^2}\left(G_0-\sqrt{G_0^2-(4\pi\beta\rho\sin\theta_0)^2}\right),
\en
where $G_0=\pi^2\rho^2+A_0^2+B_0^2$ and
\beg\label{Gapless}
\begin{split}
&A_0(\eps)=\frac{2}{\lambda_f}+P[\cos\theta_0], \\
&B_0(\eps)=P[\sin\theta_0], \\
&P[f(\eps)]=\dashint\limits_0^{\omega_D}\frac{f(\eps')\nu(\eps')d\eps'}{\eps-\eps'}.
\end{split}
\en
Here $\lambda_f$ is the value of the dimensionless coupling after the quench, $\rho(\eps)=\sqrt{\eps/\varepsilon_F}$ is the dimensionless single particle density of states and
\beg\label{eq1}
\sin\theta_0=\frac{\Delta_i}{\sqrt{\eps_\bp^2+\Delta_0^2}}, ~\cos\theta_0(\eps)=\frac{\eps_\bp}{\sqrt{(\eps_\bp^2+\Delta_0^2}}.
\en
For more details on this I would like to refer the Reader to Ref. \cite{qReview2015a}.

\end{appendix}
\end{document}